\documentstyle[12pt]{article}
\newcommand{\lD}{\hbox to0pt{\raise7pt\hbox{$\leftarrow$}\hss}D}
\newcommand{\lDelta}{\hbox to0pt{\raise7pt\hbox{$\leftarrow$}\hss}\Delta}
\newcommand{\lpa}{\hbox to0pt{\raise7pt\hbox{$\leftarrow$}\hss}\partial}
\newcommand{\slv}{\hbox{$v\kern-7pt/$}}
\newcommand{\slD}{\hbox{$D\kern-7pt/$}}
\newcommand{\slA}{\hbox{$A\kern-7pt/$}}
\newcommand{\slDa}{\hbox to0pt{\raise7pt\hbox{$\leftarrow$}\hss}\slD}

\begin{document}
\title{QCD Sum Rule for Heavy Baryons}
\author {{\small  H.Y. Jin and J.G. K\"orner
\thanks{jhy@thep.physik.uni-mainz.de,
koerner@thep.physik.uni-mainz.de}}\\
{\small Institut f\"ur Physik,
Johannes Gutenberg-Universit\"at,
Staudinger Weg 7, D 55099 Mainz , Germany }\\}
\date {}
\maketitle
\begin{center}
\begin{abstract}
We construct the heavy baryonic currents by using the Bethe-Salpeter wave 
functions in the heavy quark limit. We discuss the one-loop
renormalization of these heavy baryonic currents 
 as well as their two-point correlators up to the order 
$1/M_h$. For a special case, we do the QCD sum rule for 
masses of the doublet  (3/2,5/2). 
\end{abstract}
\end{center}
\par 

Many baryons with  spins larger than 3/2 have been found in 
the experiments. 
A successful description of these particles in theory will further 
confirm the quark model. However, up to now there is few systemic 
methods to study such kind 
of particles. On the one hand, any relativistic model 
for a three-body system will be very difficult to be dealt
with. On the other hand, there are
also some problems to apply the non-relativistic models to 
baryons, not only  because it is still difficult to get a
solution of a three-body system but also because of suspectable 
validity of the non-relativistic approximation. In order to get more 
knowledge of the excited baryons, it may be useful
to start our study from the simplest system. 
 The heavy baryons, which  contain only one heavy quark, probably 
are what we are looking for . In  such  systems, 
the spin of the heavy quark 
decouples with the light freedoms when the mass of the heavy quark  
goes to 
infinity. This is so called the heavy quark symmetry(HQS) \cite{IW}. 
Then the total spin of the heavy baryons can be constructed in
the frame of the diquark
picture. The  effects of HQS breaking can be taken into account 
perturbatively in the expansion of $1/M_h$ in the frame of the 
heavy quark effective theory(HQET) \cite{Geo}.   
Conversely, the property of
the excited heavy baryons could check whether HQET works well in 
 such a system( it is still amazing why the life time of $\Lambda_b$ is so smaller than
that of $B$).  Studying excited heavy baryons  
also provides useful information for semileptonic decays of heavy baryons 
which can serve to determine Kobayashi-Maskawa matrix
elements. The reliable 
estimate of  the widths of $\Lambda_b\rightarrow\Lambda_c^{exited}+l\nu$, 
which  probably are small, could give a  more tight constraint on 
$\Lambda_b\rightarrow\Lambda_c+l\nu$ \cite{DEL}. 

Based on HQET, a non-perturbative method is still needed to
deal with the heavy baryons.  QCD sum rule as a model-independent 
method is very powerful and has wide applications 
\cite{shifman}.  Our
discussion in this manuscript  will 
be restricted in its framework. In the following we firstly discuss
the choice of the baryonic currents in section 1. 
Then the general formulas of 
two-point correlators are discussed in section 2. In section 3
we give an example of the calculation for the masses of  the
 heavy baryon doublet (3/2,5/2). 

\section{Baryonic currents}
In the heavy quark limit, it is possible to choose  
a baryonic current that only annihilates the
heavy baryon state with the same quantum number.  
This unique character is also valid up to order $1/M_h$ in QCD sum 
rule. In th practical application, this character is  good  
enough at least for $b$ heavy baryons.  
    
There are many ways to construct such  baryonic
currents. In this paper, we would like to deduce them from the
Bethe-Salpeter wave
functions.   Let us consider the heavy baryon wave function 
\begin{equation}
\displaystyle{
\langle 0|Th(x_1)q_1(x_2)q_2(x_3)|B\rangle =e^{-im_Bv\cdot x_1}\int 
\frac{d^4p}{(2\pi)^4}
\frac{d^4k}{(2\pi)^4} 
e^{-ip\cdot x}e^{-ik\cdot x^\prime}\chi(v,p,k),} 
\end{equation}
where $x=x_2-x_3$, $x^\prime=x_1-\frac{x_2+x_3}{2}$,$v$ is the velocity of 
the baryon. We have already set $m_h\rightarrow\infty$ and suppressed 
the color index for convenience .  For an  arbitrary spin 
$n+\frac{1}{2}$, $\chi(v,p,k)$ has two independent structures \cite{bs}.
For the  parity $P=(-1)^n$,  
\begin{equation}\label{even}
\begin{array}{lll}
\chi(v,p,k) &=& u^{\mu_1\cdots\mu_j}\Phi_{\mu_1\cdots\mu_n}\gamma_5C \\
\chi(v,p,k) &=&\displaystyle{
(\sum_{S(\mu_1\cdots\mu_{n+1})}\gamma_5\gamma^{\mu_1}_\bot
u^{\mu_2\cdots\mu_{n+1}}) \Phi_{\mu_1\cdots\mu_{n+1}}C,}
\end{array}
\end{equation}
and for   $P=(-1)^{n+1}$ ,
\begin{equation}\label{odd}  
\begin{array}{lll}
\chi(v,p,k) &=& u^{\mu_1\cdots\mu_n}\Phi_{\mu_1\cdots\mu_n}C \\
\chi(v,p,k)&=&\displaystyle{
(\sum_{S(\mu_1\cdots\mu_{n+1})}\gamma_5\gamma^{\mu_1}_\bot
u^{\mu_2\cdots\mu_{n+1}}) \Phi_{\mu_1\cdots\mu_{n+1}}\gamma_5C,}     
\end{array}                                                        
\end{equation}
where $u^{\mu_1\cdots\mu_n}$ is a general Rarita-Schwinger tensor spinor, 
 $\sum_{S(\mu_1\cdots\mu_n)}$ makes indices $\mu_1\cdots\mu_n$ symmetric
and 
$\Phi_{\mu_1\cdots\mu_n}$ is a transversely  symmetric  tensor function of
$\gamma_\bot$, $p_\bot$ and $k_\bot$.   
The $\gamma_\bot$ (so as
$p_\bot$ and $k_\bot$) is 
defined as 
\begin{equation}
\gamma_\bot^\mu=\gamma^\mu-\slv v^\mu.   
\end{equation}
The first and second state of (\ref{even}) degenerate with the second
and first state of (\ref{odd}) resp. 
$n=0$ is a special case, in which the first state of (\ref{even}) is a
singlet.

Then the baryonic currents can be constructed straightly. For
instance, we can choose 
\begin{equation}\label{cur}
J_+^{\mu_1\cdots\mu_n}(x)=\Gamma_+^{\mu_1\cdots\mu_n,\nu_1\cdots\nu_n}h(x)
j_{\nu_1\cdots\nu_n}(x)
\end{equation}
for the first state of (\ref{even}), where
\begin{equation}
\Gamma_+^{\mu_1\cdots\mu_n,\nu_1\cdots\nu_n}=
g_\bot^{\mu_1\nu_1}\cdots g_\bot^{\mu_n\nu_n}
-\frac{1}{2n+1}\sum_{S(1\cdots n)}
\gamma_\bot^{\mu_1}\gamma_\bot^{\nu_1}g_\bot^{\mu_2\nu_2}
\cdots g_\bot^{\mu_n\nu_n}, 
\end{equation}
and where 
\begin{equation}\label{j}
\begin{array}{lll}
j_{\nu_1\cdots\nu_n}& =&\displaystyle{T_{\nu_1\cdots \nu_n}-\frac{1}{2n-1}
\sum_{S(\nu_1\cdots \nu_n)}g_{\bot\nu_1\nu_2}T^\nu_{\nu\nu_3\cdots \nu_n}+\cdots}\\
&+&\displaystyle{\frac{(-1)^\frac{n}{2}}{(2n-1)(2n-3)\cdots(n+1)}
\sum_{S(\nu_1\cdots \nu_n)}g_{\bot\nu_1\nu_2}\cdots
g_{\bot\nu_{n-1}\nu_n}
T^{u_1\cdots u_{\frac{n}{2}}}_{u_1\cdots u_{\frac{n}{2}}}}
\end{array}
\end{equation}
for an even $n$ and
\begin{equation}\label{j1}
\begin{array}{lll}
j_{\nu_1\cdots\nu_n}&=&\displaystyle{T_{\nu_1\cdots \nu_n}-\frac{1}{2n-1}
\sum_{S(\nu_1\cdots \nu_n)}g_{\bot\nu_1\nu_2}T^\nu_{\nu\nu_3\cdots \nu_n}+\cdots}\\
&+&\displaystyle{\frac{(-1)^\frac{n-1}{2}}{(2n-1)(2n-3)\cdots(n+2)}
\sum_{S(\nu_1\cdots \nu_n)}g_{\bot\nu_1\nu_2}\cdots
g_{\bot\nu_{n-2}\nu_{n-1}}
T^{u_1\cdots u_{\frac{n-1}{2}}}_{u_1\cdots u_{\frac{n-1}{2}}\nu_n}}
\end{array}
\end{equation}
for an odd $n$. 
In (\ref{j}) $T_{\nu_1\cdots \nu_n}$ is defined as 
$$ q_1(x)^TC\gamma_5\sum_{S(\nu_1\cdots \nu_n)}
D_{\bot\nu_1}\cdots D_{\bot\nu_{n-1}}\gamma_{\bot\nu_n}q_2(x),$$
where $g_\bot^{\mu\nu}=g^{\mu\nu}-v^\mu v^\nu$
and $D_\mu=\partial_\mu+igA_\mu$.
 It is not difficult to check 
(\ref{cur}) is orthogonal to other states  in (\ref{even}) and
(\ref{odd}). 
For instance, (\ref{cur}) is orthogonal to the second state of
(\ref{even}) and the first state of (\ref{odd}), because 
$\int d^4pd^4kTr\{C\gamma_5\Phi_{\mu_1\cdots\mu_n}C\}=0$. For the 
second state of (\ref{odd}), we can write its matrix element 
of the current (\ref{cur}) as  
\begin{equation}
\begin{array}{lll}
\langle 0|J_+^{\mu_1\cdots\mu_n}(x)|B\rangle&=&
\displaystyle{\Gamma_+^{\mu_1\cdots\mu_n\nu_1\cdots\nu_n}
(\sum_{S(\rho_1\cdots\rho_n)}\gamma_5\gamma^{\rho_1}_\bot
u^{\rho_2\cdots\rho_n})}\\
&& \displaystyle{
\sum_{S(\nu_1\cdots\nu_n)}\sum_{S(\rho_1\cdots\rho_n)}
(Ag_\bot^{\nu_1\rho_1}\cdots g_\bot^{\nu_n\rho_n}
+Bg_\bot^{\nu_1\nu_2}g_\bot^{\rho_1\rho_2}g_\bot^{\nu_3\rho_3}\cdots 
g_\bot^{\nu_n\rho_n}+\cdots)},
\end{array}
\end{equation}
where the term proportional to $A$ vanishes, the rest are also zero 
because  $\gamma_5\gamma^{\rho_1}_\bot u^{\rho_2\cdots\rho_n}$ 
vanishes after  contraction of any two indices $\rho_i$.
 
Similarly we choose 
\begin{equation}\label{cur1} 
J_-^{\mu_1\cdots\mu_n}(x)=
\Gamma_-^{\mu_1\cdots\mu_n\nu_1\cdots\nu_n}h(x)
j_{\nu_1\cdots\nu_n}(x),
\end{equation}
as the baryonic current of the second state of (\ref{odd}),
where 
\begin{equation}
\Gamma_-^{\mu_1\cdots\mu_n\nu_1\cdots\nu_n}=
\sum_{S(i\cdots n)}
\gamma_\bot^{\mu_1}\gamma_\bot^{\nu_1}g_\bot^{\mu_2\nu_2}
\cdots g_\bot^{\mu_n\nu_n}.
\end{equation}
The currents for the rest states can be obtained by omitting 
the $\gamma_5$ inserted in  the two light quarks.
 It does not change the
quantum number (except isospin)  of the currents to  
insert  $\slv$ and $\slD$ into $q^Tq$, or replace   insertion of
$\gamma_\bot^\nu$ by $D_\bot^\nu$, or  let $D_\bot^\mu$ act on $q^T$.
But these insertions probably affects the isospin character. 
For instance, in the case of $n=1$, currents (\ref{cur}) and 
(\ref{cur1}) only have $I=0$ state. After the insertion of $\slv$, 
they only have $I=1$ state. 
We will go back to discuss this later. Giving a definite quantum number, we only consider the currents with the lowest dimension, because
 higher dimension currents make QCDSR more unstable.   
So we will not
discuss the currents with  insertion of $\slD$ later. 

The baryonic currents also can be obtained by using 
the wave functions proposed in \cite{falk}. The result is the
same as above.    

Now we are going on  the renormalization of these baryonic
currents, which must be used in the QCD sum rule beyond $\alpha_s$ 
correction as well as matching the matrix elements of these currents 
from HQET to the full QCD. 
The one-loop renormalization of these baryonic currents
without the covariant derivative $D_\mu$ has  been considered in 
\cite{grozin}. When we consider the renormalization of 
 the currents with one or more derivative $D_\mu$, 
the currents with 
the same quantum number and dimension will mix with each other.   
 For convenience we firstly consider a symmetric tensor current,
 which is transverse, traceless and  only has
one covariant derivative $D_\mu$, i.e. 
\begin{equation}\label{j2} 
j_1^{\mu\nu}=q_1^TC\Gamma(D^\mu_\bot\gamma^\nu_\bot+D^\mu_\bot
\gamma^\nu_\bot-\frac{2}{3}g^{\mu\nu}_\bot\slD)q_2 h.  
\end{equation}
Obviously the renormalization of $j_1^{\mu\nu}$ is the same as 
$J_\pm^{\mu\nu}$ in (\ref{cur}) and (\ref{cur1}) in the heavy 
quark limit. The complete set of operators which may mix with (\ref{j2}) is
\begin{equation}
\begin{array}{lll} 
j_1^{\mu\nu}&=&q_1^TC\Gamma(D^\mu_\bot\gamma^\nu_\bot+D^\nu_\bot
\gamma^\mu_\bot-\frac{2}{3}g^{\mu\nu}_\bot\slD)q_2 h\\
j_2^{\mu\nu}&=&q_1^TC\Gamma(\lD^\mu_\bot\gamma^\nu_\bot+\lD^\nu_\bot
\gamma^\mu_\bot-\frac{2}{3}g^{\mu\nu}_\bot\slDa_\bot)q_2 h\\
j_3^{\mu\nu}&=&q_1^TC\Gamma(\gamma^\nu_\bot q_2 D^\mu_\bot h
+\gamma^\mu_\bot q_2 D^\nu_\bot h
-\frac{2}{3}g^{\mu\nu}_\bot\gamma_{\bot\rho} q_2 D^\rho_\bot h),\\
\end{array}
\end{equation}
where $\Gamma$ could be $1$ $\gamma_5$, $\slv$ or  $\gamma_5\slv$ and 
$q^T\lD^\mu_\bot=(D^\mu_\bot q)^T$. $j_3^{\mu\nu}$ is trivial because
it differs from  $j_1^{\mu\nu}+j_2^{\mu\nu}$ by a total
derivative. The 
nuisance operators such as 
$$ q_1^TC\Gamma(A^\mu_\bot\gamma^\nu_\bot+A^\nu_\bot
\gamma^\mu_\bot-\frac{2}{3}g^{\mu\nu}_\bot\slA)q_2 h$$ 
do not appear because they do not vanish by the equations of the motion. 

Then the bare current $[j_1^{\mu\nu}]$ can be expressed as 
\begin{equation}
[j_1^{\mu\nu}]=Z_qZ_h^{\frac{1}{2}}
(Z_1\tilde j_1^{\mu\nu}+Z_2\tilde j_2^{\mu\nu}+Z_3\tilde j_3^{\mu\nu}),
\end{equation}
where $Z_q=1-C_F\frac{\alpha_s}{4\pi\epsilon}$ and 
$Z_h=1+2C_F\frac{\alpha_s}{4\pi\epsilon}$ are the
renormalization coefficients of 
the light quark and the heavy quark resp. and 
$\tilde j_i^{\mu\nu}(i=1,3)$ are the renormalized currents.
 In order to determine the renormalization coefficients
$Z_i(i=1,3)$, we need to insert the current into 1PI 
diagrams and extract the ultraviolet divergence. We find that the 
divergence associated with the three-quark vertex are sufficient 
to determine all counterterms. The relevant Feynman diagrams are shown 
in Fig.1. In the  Feynman gauge, we obtain 
\begin{equation}\label{rj2}
\begin{array}{lll}
Z_1&=&1+[(\frac{H^2}{6}+3/2)C_B-C_F]\frac{\alpha_s}{4\pi\epsilon}\\
Z_2&=&(\frac{H^2}{12}+1)C_B\frac{\alpha_s}{4\pi\epsilon}\\
Z_3&=&0,
\end{array}
\end{equation}
where $\gamma_\mu\Gamma\gamma^\nu_\bot\gamma^\mu=H\Gamma\gamma^\nu_\bot$,
$C_B=\frac{N_c+1}{2N_c}$,$C_F=\frac{N_c^2-1}{2N_c}$ and 
$\epsilon$ is defined as $D=4-2\epsilon$ in the 
$\rm\overline{MS}$-scheme. The renormalization of $j_2^{\mu\nu}$ is 
similar as that of $j_1^{\mu\nu}$. Therefore  $j_1^{\mu\nu}$ and 
$j_2^{\mu\nu}$ can construct a renormalization-invariant space. 
The one-loop renormalization invariant currents can be chosen as 
 $$j_\pm^{\mu\nu}=j_1^{\mu\nu}\pm j_2^{\mu\nu}.$$  
They satisfy the equation 
\begin{equation}
[j_\pm^{\mu\nu}]= Z_qZ_h^{\frac{1}{2}}(Z_1\pm Z_2)\tilde j_\pm^{\mu\nu}.
\end{equation}
The current $j_+^{\mu\nu}$ corresponds to the state $I=1$ for 
$\Gamma=1$, $\slv$ or $\gamma_5\slv$ or $I=0$ for $\Gamma=\gamma_5$. 
$j_-^{\mu\nu}$ corresponds to  $I=0$ state for $\Gamma=1$, $\slv$, 
or $\gamma_5\slv$ or $I=1$ for $\Gamma=\gamma_5$.  

The anomalous dimensions are obtained by 
\begin{equation}
\begin{array}{lll}
\gamma_\pm &=& \displaystyle{\frac{d log(Z_1\pm Z_2)}{d log\mu}}\\
&=& \displaystyle{
-\left [(\frac{H^2}{6}+3/2)C_B-C_F\pm(\frac{H^2}{12}+1)C_B \right ]
\frac{\alpha_s}{2\pi}},
\end{array}
\end{equation}
where $\mu$ is the renormalization  point. The relation between HQET´s 
currents and the corresponding ones in  the full QCD in the leading 
logarithmic approximation is 
\begin{equation}
J_{QCD} = C(\mu)J_{HQET}(\mu),
\end{equation} 
where the matching condition is 
\begin{equation}
C_\pm(\mu)=C_\pm(M_h)
\displaystyle{e^{\int^{g(\mu)}_{g(M_h)}dg\frac{\gamma_\pm(g)}{\beta(g)}}}
\end{equation}
and where $C_\pm(M_h)=1+{\cal{O}}(\alpha_s(M_h))$. 
It is more complicated to get the full matching condition. The relevant 
references could be seen in \cite{E}\cite{grozin}. We do not discuss this
case. 

(\ref{rj2}) can be extended to a general case. 
For the rank $n$ tensor current with $n-1$ covariant derivative $D$
, the complete set of operators is 
\begin{equation}\label{nrank}
\begin{array}{lll}
j^i_{\nu_1\cdots\nu_n}& =&\displaystyle{(T^i_{\nu_1\cdots \nu_n}-\frac{1}{2n-1}
\sum_{S(\nu_1\cdots \nu_n)}g_{\bot\nu_1\nu_2}T^{i~\nu}_{\nu\nu_3\cdots \nu_n}+\cdots}\\
&+&\displaystyle{\frac{(-1)^\frac{n}{2}}{(2n-1)(2n-3)\cdots(n+1)}
\sum_{S(\nu_1\cdots \nu_n)}g_{\bot\nu_1\nu_2}\cdots
g_{\bot\nu_{n-1}\nu_n}
T^{i~u_1\cdots u_{\frac{n}{2}}}_{u_1\cdots u_{\frac{n}{2}}})h(i=1,n)}
\end{array}
\end{equation}
for an even $n$ and
\begin{equation}\label{nrank1}
\begin{array}{lll}
j^i_{\nu_1\cdots\nu_n}&=&\displaystyle{(T^i_{\nu_1\cdots \nu_n}-\frac{1}{2n-1}
\sum_{S(\nu_1\cdots \nu_n)}g_{\bot\nu_1\nu_2}T^{i~\nu}_{\nu\nu_3\cdots \nu_n}+\cdots}\\
&+&\displaystyle{\frac{(-1)^\frac{n-1}{2}}{(2n-1)(2n-3)\cdots(n+2)}
\sum_{S(\nu_1\cdots \nu_n)}g_{\bot\nu_1\nu_2}\cdots
g_{\bot\nu_{n-2}\nu_{n-1}}
T^{i~u_1\cdots u_{\frac{n-1}{2}}}_{u_1\cdots u_{\frac{n-1}{2}}\nu_n})
h(i=1,n)}
\end{array}
\end{equation}
for an odd $n$,
where 
\begin{equation}\label{T}
T^i_{\mu_1\cdots \mu_n}=q_1(x)^TC\Gamma\sum_{S(\nu_1\cdots \nu_n)} 
\lD_{\bot\nu_1}\cdots\lD_{\bot\nu_{i-1}}
D_{\bot\nu_i}\cdots D_{\bot\nu_{n-1}}\gamma_{\bot\nu_n}q_2(x).
\end{equation}
The bare current $[j_i^{\mu_1\cdots\mu_n}]$ could be written as 
\begin{equation}
[j_i^{\mu_1\cdots\mu_n}]=Z_qZ_h^{\frac{1}{2}}\sum_{k=1,n}
Z_{ik}{\tilde j}_k^{\mu_1\cdots\mu_n}.
\end{equation}
After calculating the Feynman diagrams similar to Fig.1, we obtain  
\begin{equation}
\begin{array}{lll}
Z_{ik}&=&\displaystyle{1+\left[\right 
 .\{(\frac{1}{n-k+1}+\frac{1}{k})C_B-(\sum_{m=0}^{n-k-1}\frac{m+1}
{(n-k+1)(n-k-m)}+\sum_{m=0}^{k-2}\frac{m+1}{k(k-m-1)})2C_F\}}\\
&&\displaystyle{
+\sum_{m=0}^{k-1}\sum_{l=0}^mC_{k-1}^{m}C_{n-k}^{k-m-1}C_m^l
\frac{(m-l)!(n-k+1)!(k-l-1)!^2}{k!(n-l+1)!}\frac{H^2}{2}C_B\left .\right
]\frac{\alpha_s}{4\pi\epsilon}}
\end{array}
\end{equation}
for $i= k$, 
\begin{equation}
\begin{array}{lll}
Z_{ik}&=&\displaystyle{\left [\right .
\sum_{m=0}^{i-1}\sum_{l=0}^mC_{k-1}^{m}C_{n-k}^{i-m-1}C_m^l
\frac{(m-l)!(i-l-1)!(n-k+1)!(k-l-1)!}{i!(n-l+1)!}\frac{H^2}{2}C_B}\\
&&\displaystyle{
+C_{k-1}^{i-1}\frac{(k-i-1)!(n-k+1)!}{(n-i+1)!}2C_B\left .\right ]
\frac{\alpha_s}{4\pi\epsilon}}
\end{array}
\end{equation}
for $i< k$ and
\begin{equation}
\begin{array}{lll}
Z_{ik}&=&\displaystyle{\left[\right 
. \sum_{m=0}^{k-1}\sum_{l=0}^mC_{k-1}^{m}C_{n-k}^{i-m-1}C_m^l
\frac{(m-l)!(i-l-1)!(n-k+1)!(k-l-1)!}{i!(n-l+1)!}\frac{H^2}{2}C_B}\\
&&\displaystyle{
+C_{n-k}^{n-i}\frac{(i-k-1)!k!}{i!}2C_B\left .\right ]
\frac{\alpha_s}{4\pi\epsilon}}
\end{array}
\end{equation}
for $i> k$. The renormalization-invariant currents can be obtained 
through the diagonization of this matrix. 
We do not find a simple way to diagonize it. However,
it should not be difficult to  get the numerical solution  
for a special $n$.  

\section{Two-point correlator}
The basic idea of the QCD sum rule is to consider  the two-point
correlator
\begin{equation}\label{cor}
\Pi_{\mu_1\cdots\mu_n,\nu_1\cdots\nu_n}(k)=\int d^4x e^{ik\cdot x}
i\langle
0|J^i_{\mu_1\cdots\mu_n}(x)\bar J^j_{\nu_1\cdots\nu_n}|0\rangle
\end{equation}
for a region of $k$  in which one can incorporate the asymptotic freedom
property of QCD via the operator product expansion (OPE), and then 
relate it to the hardonic matrix elements via the dispersion relation. The currents $J^i_{\mu_1\cdots\mu_n}(x)$ and 
$J^i_{\mu_1\cdots\mu_n}(x)$ were defined in (\ref{cur})
or (\ref{cur1}). They have the same quantum number and the same
$\Gamma$ defined in (\ref{T}). 
We do not discuss the case that these two currents have 
different  $\Gamma$, because the perturbation part of 
such correlators  is zero.  The extra indices 
$i,j$ were defined in (\ref{T}).   

In order to analyze the construction of the correlator 
(\ref{cor}) up to the order $1/M_h$, we write down  
the effect Lagrangian of the heavy quark  up to the order $1/M_h$  
\begin{equation}
L_{eff}= \bar h_viv\cdot Dh_v+\frac{1}{2M_h}(\bar h_v(iD)^2h_v
-\frac{1}{2}\bar h_vgG_{\mu\nu}\sigma_\bot^{\mu\nu}h_v).
\end{equation}
In the leading order of HQET, the the spin of the heavy quark 
decouples with the light freedom, therefore (\ref{cor}) can be 
expressed as
\begin{equation}\label{le}  
\Pi_0^{\mu_1\cdots\mu_n,\nu_1\cdots\nu_n}(\omega)=\Pi_0(\omega)
I_{\alpha_1\cdots\alpha_n,\beta_1\cdots\beta_n}
\Gamma_{\pm}^{\mu_1\cdots\mu_n,\alpha_1\cdots\alpha_n}
\frac{1+\slv}{2}\bar\Gamma_{\pm}^{\nu_1\cdots\nu_n,\beta_1\cdots\beta_n},
\end{equation}
where $I_{\alpha_1\cdots\alpha_n,\beta_1\cdots\beta_n}$ 
is transverse, symmetric and traceless for indices
$\alpha_1\cdots\alpha_n$ and 
$\beta_1\cdots\beta_n$ resp.. 
Since $x_\mu$ is proportional to $v_\mu$ 
via the heavy quark propagator $\frac{1+\slv}{2}\int dt \delta(x-vt)$,
only one scalar argument $\omega=k\cdot v$ survives in (\ref{le}). 
 $I_{\alpha_1\cdots\alpha_n,\beta_1\cdots\beta_n}$ is only 
composed of $v_\mu$ and $ g_{\mu\nu}$ and  
must be written as 
\begin{equation}
\begin{array}{lll}
I_{\alpha_1\cdots\alpha_n,\beta_1\cdots\beta_n}&=&
\displaystyle{
\sum_{S(\alpha_1\cdots\alpha_n)}
\sum_{S(\beta_1\cdots\beta_n)}
(g_{\bot\alpha_1\beta_1}\cdots
g_{\bot\alpha_n\beta_n}-\frac{2}{2n-1}g_{\bot\alpha_1\alpha_2}
g_{\bot\beta_1\beta_2}g_{\bot\alpha_3\beta_3}\cdots 
g_{\bot\alpha_n\beta_n}}\\
&&
\displaystyle{
+\cdots +
\frac{(-2)^{\frac{n}{2}}}{(2n-1)(2n-3)\cdots (n+1)}
g_{\bot\alpha_1\alpha_2}g_{\bot\beta_1\beta_2}
\cdots g_{\bot\alpha_{n-1}\alpha_n}g_{\bot\beta_{n-1}\beta_n})}
\end{array}
\end{equation}
for an even $n$ and 
\begin{equation}
\begin{array}{lll}
I_{\alpha_1\cdots\alpha_n,\beta_1\cdots\beta_n}&=&
\displaystyle{\sum_{S(\alpha_1\cdots\alpha_n,)}
\sum_{S(\beta_1\cdots\beta_n,)}
(g_{\bot\alpha_1\beta_1}\cdots
g_{\bot\alpha_n\beta_n}-\frac{2}{2n-1}g_{\bot\alpha_1\alpha_2}
g_{\bot\beta_1\beta_2}g_{\bot\alpha_3\beta_3}\cdots
g_{\bot\alpha_n\beta_n}}\\
&&
\displaystyle{+\cdots +
\frac{(-2)^{\frac{n-1}{2}}}{(2n-1)(2n-3)\cdots (n+2)}
g_{\bot\alpha_1\alpha_2}g_{\bot\beta_1\beta_2}
\cdots g_{\bot\alpha_{n-2}\alpha_{n-1}}g_{\bot\beta_{n-2}\beta_{n-1}}
g_{\bot\alpha_n\beta_n})}
\end{array}
\end{equation}
for an odd $n$. 

To the order of $1/M_h$, the term $\frac{1}{2M_h}
\bar h_v(iD)^2h_v$ does not break the spin symmetry but could produce a new scalar argument $k^2$ in the correlator (\ref{cor}).  
For convenience, 
we choose k paralleling to $v$. Then we also can
write its contribution as
\begin{equation}\label{1a} 
\Pi_{1a}^{\mu_1\cdots\mu_n,\nu_1\cdots\nu_n}(\omega)=
\frac{1}{2M_h}\Pi_{1a}(\omega)
I_{\alpha_1\cdots\alpha_n,\beta_1\cdots\beta_n}
\Gamma_{\pm}^{\mu_1\cdots\mu_n,\alpha_1\cdots\alpha_n}
\frac{1+\slv}{2}\bar\Gamma_\pm^{\nu_1\cdots\nu_n,\beta_1\cdots\beta_n}.
\end{equation}
The chromo-magnetic term -$\frac{1}{2M_h}\frac{1}{2}\bar
h_vgG_{\mu\nu}\sigma_\bot^{\mu\nu}h_v$ breaks spin symmetry and 
causes the mass-split of the heavy baryon doublets. 
We can write its contribution as  
\begin{equation}\label{1b}
\Pi_{1b}^{\mu_1\cdots\mu_n,\nu_1\cdots\nu_n}(\omega)=
\frac{1}{2M_h}i\Pi_{1b}(\omega)
\Gamma_{\pm}^{\mu_1\cdots\mu_n,\alpha_1\cdots\alpha_n}
\frac{1+\slv}{2}\sigma_\bot^{ij}
F_{ij,\alpha_1\cdots\alpha_n,\beta_1\cdots\beta_n}
\frac{1+\slv}{2}\bar\Gamma_{\pm}^{\nu_1\cdots\nu_n,\beta_1\cdots\beta_n},
\end{equation}
where   $F_{ij,\alpha_1\cdots\alpha_n,\beta_1\cdots\beta_n}$
is transverse, symmetric and traceless for indices
$\alpha_1\cdots\alpha_n$ and $\beta_1\cdots\beta_n$ resp.. Since 
$F_{ij,\alpha_1\cdots\alpha_n,\beta_1\cdots\beta_n}$ also 
only depends on $v_\mu$ and $g_{\mu\nu}$, we can write down 
\begin{equation}
\begin{array}{lll}
\sigma_{\bot}^{ij}F_{ij,\alpha_1\cdots\alpha_n,\beta_1\cdots\beta_n}&=&
\displaystyle{\sum_{S(\alpha_1\cdots\alpha_n)}
\sum_{S(\beta_1\cdots\beta_n)}
\left (\right . g_{\bot\alpha_1\beta_1}\cdots
g_{\bot\alpha_{n-1}\beta_{n-1}}}\\
&&\displaystyle{
-\frac{2}{2n-1}g_{\bot\alpha_1\alpha_2}
g_{\bot\beta_1\beta_2}g_{\bot\alpha_3\beta_3}\cdots
g_{\bot\alpha_{n-1}\beta_{n-1}}+\cdots +}\\
&&\displaystyle{
\frac{(-2)^{\frac{n-1}{2}}}{(2n-1)(2n-3)\cdots (n+2)}
g_{\bot\alpha_1\alpha_2}g_{\bot\beta_1\beta_2}
\cdots
g_{\bot\alpha_{n-2}\alpha_{n-1}}g_{\bot\beta_{n-2}\beta_{n-1}}\left .
\right )
\sigma_{\bot\alpha_n\beta_n}}
\end{array}
\end{equation}
for an odd $n$ and 
\begin{equation}
\begin{array}{lll}
\sigma_{\bot}^{ij}F_{ij,\alpha_1\cdots\alpha_n,\beta_1\cdots\beta_n} &=&
\displaystyle{
\sum_{S(\alpha_1\cdots\alpha_n,)}
\sum_{S(\beta_1\cdots\beta_n,)}
\left (g_{\bot\alpha_1\beta_1}\cdots
g_{\bot\alpha_{n-2}\beta_{n-2}}\right .}\\
&&\displaystyle{
-\frac{2}{2n-1}g_{\bot\alpha_1\alpha_2}
g_{\bot\beta_1\beta_2}g_{\bot\alpha_3\beta_3}\cdots
g_{\bot\alpha_{n-2}\beta_{n-2}}+\cdots +}\\
&&\displaystyle{
\frac{(-2)^{\frac{n-2}{2}}}{(2n-1)(2n-3)\cdots (n+3)}
g_{\bot\alpha_1\alpha_2}g_{\bot\beta_1\beta_2}
\cdots g_{\bot\alpha_{n-3}\alpha_{n-2}}g_{\bot\beta_{n-3}\beta_{n-2}}\left
.\right )}\\
&& g_{\bot\alpha_{n-1}\beta_{n-1}}\sigma_{\bot\alpha_n\beta_n}
\end{array}
\end{equation}
for an even $n$.

It is not difficult to check that 
\begin{equation}\label{qm}
\begin{array}{lll}
\Gamma_{+}^{\mu_1\cdots\mu_n,\alpha_1\cdots\alpha_n}
\frac{1+\slv}{2}\sigma_\bot^{ij}
F_{ij,\alpha_1\cdots\alpha_n,\beta_1\cdots\beta_n}
&=&
-inI_{\alpha_1\cdots\alpha_n,\beta_1\cdots\beta_n}
\Gamma_{+}^{\mu_1\cdots\mu_n,\alpha_1\cdots\alpha_n}
\frac{1+\slv}{2},\\
\Gamma_{-}^{\mu_1\cdots\mu_n,\alpha_1\cdots\alpha_n}
\frac{1+\slv}{2}\sigma_\bot^{ij}
F_{ij,\alpha_1\cdots\alpha_n,\beta_1\cdots\beta_n}\\
&=& 
i(n+1)I_{\alpha_1\cdots\alpha_n,\beta_1\cdots\beta_n}
\Gamma_{-}^{\mu_1\cdots\mu_n,\alpha_1\cdots\alpha_n}
\frac{1+\slv}{2}.
\end{array}
\end{equation}
This  is consistent with the naive expectation based on the quantum 
mechanics.

With these formulas in our hand, we begin to construct the QCD sum 
rule for masses of the heavy baryons. 
At first, the imaginary part (\ref{cor}) can be obtained by
insertion of  the physical states. 
On the assumption that only one resonance and continuum 
contribute to the correlator (\ref{cor}), we can write  
\begin{equation}\label{phe}
\begin{array}{lll}
Im\Pi_{\mu_1\cdots\mu_n,\nu_1\cdots\nu_n}(\omega)&=&
\int d^4x e^{ik\cdot x} \langle
0|J_{\mu_1\cdots\mu_n}(x)|B_{n\pm\frac{1}{2}}\rangle
\langle B_{n\pm\frac{1}{2}}|J_{\nu_1\cdots\nu_n}|0\rangle + continnum\\
&=&\pi\delta(\omega-\Lambda)
|f_{n\pm\frac{1}{2}}|^2 M_{n\pm\frac{1}{2}}^{2n+3}
I_{\alpha_1\cdots\alpha_n,\beta_1\cdots\beta_n}
\Gamma_{\pm}^{\mu_1\cdots\mu_n,\alpha_1\cdots\alpha_n}
\frac{1+\slv}{2}
\bar\Gamma_{\pm}^{\beta_1\cdots\beta_n,\nu_1\cdots\nu_n}\\
&& + continuum,
\end{array}
\end{equation}
where $M_{n\pm\frac{1}{2}}$ is the mass of the heavy baryon,
 $\Lambda$ is defined as
$$M_{n\pm\frac{1}{2}}=M_h+\Lambda+{\cal{O}}(\frac{1}{M^2_h}) $$ 
and 
\begin{equation}\label{ff}
\langle 0|J_{\mu_1\cdots\mu_n}|B_{n\pm\frac{1}{2}}(k)\rangle
=if_{n\pm\frac{1}{2}}M_{n\pm\frac{1}{2}}^{n+\frac{3}{2}}
\Gamma_{\pm}^{\mu_1\cdots\mu_n,\alpha_1\cdots\alpha_n}
\epsilon_{\alpha_1\cdots\alpha_n}u.
\end{equation}
In (\ref{ff})  $u$ is a $\frac{1}{2}$ spinor and
$\epsilon_{\alpha_1\cdots\alpha_n}$ is a 
symmetric polarization tensor and satisfies 
\begin{equation}
v^{\alpha_1}\epsilon_{\alpha_1\cdots\alpha_n}=0,~~~~~~~~~~ 
\epsilon^{\alpha}_{\alpha\alpha_3\cdots\alpha_n}=0.
\end{equation} 
$\Gamma_{\pm}^{\mu_1\cdots\mu_n,\alpha_1\cdots\alpha_n}
\epsilon_{\alpha_1\cdots\alpha_n}u$ can serve as a $n\pm\frac{1}{2}$
spinor\cite{falk}. There are $2n+1$ independent
$\epsilon_{\alpha_1\cdots\alpha_n}$ which are  normalized as
\begin{equation}
\displaystyle{
\sum_{i=1}^{2n+1}\epsilon^i_{\alpha_1\cdots\alpha_n}\epsilon^{*i}_{\beta_1\cdots\beta_n}
=I_{\alpha_1\cdots\alpha_n,\beta_1\cdots\beta_n}.}
\end{equation}

Then the sum rule can be established. The correlator (\ref{cor}) 
obey the dispersion relation
\begin{equation}\label{dis}
\Pi_\pm(\omega)=\displaystyle{\frac{1}{\pi}\int
\frac{Im\Pi_\pm(\omega^\prime) 
d\omega^\prime}{\omega^\prime-\omega}
=\int \frac{\rho_\pm(\omega^\prime)d\omega^\prime}
{\omega^\prime-\omega}},
\end{equation}
where the spectral density $\rho_\pm(\omega)=\frac{1}{\pi}
Im\Pi_\pm(\omega)$ can be obtained from (\ref{le}), (\ref{1a}), 
(\ref{1b}) and (\ref{qm}) (via OPE in a proper region of $\omega$)
 on the one hand,  
\begin{equation}
\begin{array}{l}
\Pi_\pm(\omega)=\Pi_0(\omega)+
\frac{1}{2M_h}(\Pi_{ia}(\omega)+a_\pm\Pi_{ib}(\omega)),\\
a_+=n,~~~~~~~~~~ a_-=-(n+1).
\end{array}
\end{equation}
On the other hand,
$$\rho_\pm(\omega^\prime)=\pi\delta(\omega-\Lambda)
|f_{n\pm\frac{1}{2}}|^2 M_{n\pm\frac{1}{2}}^{2n+3}+continuum$$ is
obtained from (\ref{phe}). 
Thus, the baryonic mass as well as its matrix element can be 
extracted from (\ref{dis}). One may note that there are two sources 
of 
uncertainty in (\ref{dis}). At first, we have to truncate the series of OPE. Thus, the contributions from higher dimension operators raise the main uncertainty of QCDSR. Another uncertainty is from 
the contributions of heavier resonances and continuum.  
 In order to suppress these uncertainties, the  Borel transformation is applied 
on the both side of (\ref{dis}). Then  we obtain 
\begin{equation}\label{sr}    
\displaystyle{\int^s_0 d\omega e^{-\frac{\omega}{M}}\rho_\pm(\omega)=
f_{n\pm\frac{1}{2}}f^*_{n\pm\frac{1}{2}}
M_{n\pm\frac{1}{2}}^{2n+3}e^{-\frac{\Lambda_\pm}{M}}},
\end{equation}
where $M$ is the Borel parameter and the upper bound of the integral 
$s$ is used to remove the rest contributions from the continuum. 
The mass parameter $\Lambda$ can be obtained by taking the partial derivative respect to $1/M$ on both sides of (\ref{sr})
\begin{equation}\label{mass}
\Lambda_\pm=\displaystyle{\frac{\int^s_0 d\omega
\omega\rho_\pm(\omega)e^{-\frac{\omega}{M}}}{\int^s_0 d\omega
\rho_\pm(\omega)e^{-\frac{\omega}{M}}}}.
\end{equation} 
The Borel parameter should be chosen in a region that $\Lambda_\pm$
is not sensitive to it. In  graphic language, there should be a 
plateau in the plot of M-$\Lambda_\pm$. 

\section{Sum rule for (3/2,5/2)}
For a special case, let us consider the heavy baryon doublet 
(3/2,5/2). As a preliminary calculation,  the radiative corrections are not taken in account. So the 
heavy quark propagator to the order $1/M_h$  in the coordinate
 space  can be expressed as 
\begin{equation}
S_h(x)=\frac{1+\slv}{2}\int^\infty_0 dt\left [1+\frac{it}{2M_h}
(-\partial^2-2gG_{\mu\nu}x^{\mu}\partial^\nu+\frac{\pi x^2}{24}
\langle\alpha_s G^2\rangle-\frac{gG_{\mu\nu}}{2}
\sigma_\bot^{\mu\nu}\frac{1+\slv}{2})\right ]\delta(x-vt),
\end{equation}
where we have used the fixed point gauge $x^\mu A_\mu(x)=0$. 
The two-point correlators then  are  
\begin{equation}\label{ll}
\begin{array}{ll}
\Pi_\pm^{\mu_1\mu_2\nu_1\nu_2}(x)=&-i
\left [\right . Tr\{\Gamma\Delta_{\alpha_1\alpha_2}(x)
S_l(x)\lDelta_{\beta_1\beta_2}(0)\bar\Gamma S^c_l(x)\}\\
&+(-1)^{I+1} Tr\{S^c_l(x)\lDelta_{\alpha_1\alpha_2}(x)\Gamma^c
S_l(x)\lDelta_{\beta_1\beta_2}(0)\bar\Gamma\}\left .\right ]
\Gamma_\pm^{\mu_1\mu_2,\alpha_1\alpha_2}
S_h(x)\bar\Gamma_\pm^{\nu_1\nu_2,\beta_1\beta_2},
\end{array}
\end{equation}
where 
$$\Delta^{\alpha_1\alpha_2}(x)=D_\bot^{\alpha_1}(x)\gamma_\bot^{\alpha_2}
+D_\bot^{\alpha_2}(x)\gamma_\bot^{\alpha_1}-\frac{2}{3}g_\bot^{\alpha_1\alpha_2}
\slD_\bot(x), $$
$\bar\Gamma=\gamma_0\Gamma^\dagger\gamma_0$, 
$\Gamma^c=C^{-1}\Gamma^TC$, $S^c_l(x)=C^{-1}S^T_l(x)C$, 
 I is the isospin of the current     
and the light propagator $S_l(x)$ in external gluonic field  was  
given in \cite{shifman}. The correlators in momentum space related 
to (\ref{ll}) via Fourier Transformation  
\begin{equation}\label{pa1}
\Pi_\pm^{\mu_1\mu_2\nu_1\nu_2}(k^2,\omega)=\int d^4x e^{ik\cdot x}
\Pi_\pm^{\mu_1\mu_2\nu_1\nu_2}(x) 
\end{equation}
Because we set $k_\mu$ paralleling to $v_\mu$, by using the 
technique of the partial integral, we can write (\ref{pa1}) in the 
form 
\begin{equation}\label{pa}
\begin{array}{lll}
\Pi_\pm^{\mu_1\mu_2\nu_1\nu_2}(\omega)&=&\int d^4x e^{i\omega t}
\Pi_\pm^{\prime\mu_1\mu_2\nu_1\nu_2}(t)\\
&&=\Gamma_\pm^{\mu_1\mu_2,\alpha_1\alpha_2}
I_{\alpha_1\alpha_2,\beta_1\beta_2}
 \Gamma_\pm^{\nu_1\nu_2,\beta_1\beta_2}
\int d^4x e^{i\omega t}\Pi_\pm(t)
\end{array}
\end{equation}
We analytically continue correlators from $t>0$ to imaginary 
$t=-i\tau$. Then the spectral density $\rho_\pm(\omega)$ are 
related to $\Pi_\pm(\tau)$ by the Laplace transformation 
\cite{grozin1} 
\begin{equation}
\rho_\pm(\omega)=\displaystyle{\frac{i}{2\pi}\int^{a+i\infty}_
{a-i\infty}d\tau e^{\omega\tau}\Pi_\pm(\tau)}    
\end{equation}

Fig.2 shows the Feynman diagrams that we calculate.   
The corresponding results are read as   
\begin{equation}\label{I}   
\begin{array}{lll}   
\rho_0(\omega)&=&\displaystyle{\frac{\omega^7}{630\pi^4}-
\frac{17\omega^3}{864\pi^3}\langle\alpha_sG^2\rangle}, \\  
\rho_{1a}(\omega)&=&\displaystyle{-\frac{2\omega^8}{315\pi^4}+
\frac{\omega^4}{18\pi^3}\langle\alpha_sG^2\rangle},\\
\rho_{1b}(\omega)&=&\displaystyle{-\frac{\omega^4}{216\pi^3}
\langle\alpha_sG^2\rangle}
\end{array}   
\end{equation}
for $\Gamma=1$ with $I=0$ and $\Gamma=\gamma_5$ with I=1, 

\begin{equation}\label{I1}
\begin{array}{lll}
\rho_0(\omega)&=&\displaystyle{\frac{\omega^7}{630\pi^4}-
\frac{7\omega^3}{432\pi^3}\langle\alpha_sG^2\rangle}, \\
\rho_{1a}(\omega)&=&\displaystyle{-\frac{\omega^8}{140\pi^4}+
\frac{25\omega^4}{432\pi^3}\langle\alpha_sG^2\rangle},
\\
\rho_{1b}(\omega)&=&\displaystyle{-\frac{\omega^4}{216\pi^3}
\langle\alpha_sG^2\rangle}
\end{array}  
\end{equation}

for $\Gamma=\slv$ and $\Gamma=\gamma_5\slv$ with I=0,

\begin{equation}\label{I3}   
\begin{array}{lll}
\rho_0(\omega)&=&\displaystyle{\frac{\omega^7}{630\pi^4}+
\frac{\omega^3}{864\pi^3}\langle\alpha_sG^2\rangle}, \\
\rho_{1a}(\omega)&=&\displaystyle{-\frac{\omega^8}{105\pi^4}-
\frac{\omega^4}{24\pi^3}\langle\alpha_sG^2\rangle},\\
\rho_{1b}(\omega)&=&\displaystyle{-\frac{\omega^4}{216\pi^3}
\langle\alpha_sG^2\rangle}
\end{array}
\end{equation}

for  $\Gamma=\gamma_5$ with I=0 and $\Gamma=1$ with $I=1$, 

\begin{equation}\label{I4}   
\begin{array}{lll}
\rho_0(\omega)&=&\displaystyle{\frac{\omega^7}{630\pi^4}+
\frac{\omega^3}{432\pi^3}\langle\alpha_sG^2\rangle}, \\
\rho_{1a}(\omega)&=&\displaystyle{-\frac{13\omega^8}{1260\pi^4}-
\frac{\omega^4}{54\pi^3}\langle\alpha_sG^2\rangle},\\
\rho_{1b}(\omega)&=&\displaystyle{-\frac{\omega^4}{216\pi^3}
\langle\alpha_sG^2\rangle}
\end{array}
\end{equation}

for  $\Gamma=$\slv$, \gamma_5\slv$ with I=1.

It is interesting to compare (\ref{I}) with (\ref{I1}). 
(\ref{I}) shows that $I=0$ state of the current 
with  $\Gamma=\slv$ and $\Gamma=\Gamma_5\slv$ are degenerate 
up to the order we consider. However, from (\ref{I1}),  
$I=0$ state of the current with $\Gamma=1$ is degenerate with  
$I=1$ state of the current with $\Gamma=\Gamma_5$ (not naively expected 
I=0 state).  It is another example that the insertion of $\slv$ 
affects isospin.  The degeneration of (\ref{I})-(\ref{I4}) 
will disappear when the light quark mass or the quark condensate is 
taken into account, i.e., the chiral symmetry is broken.   
Since our  calculation in this paper is very preliminary, we leave 
the calculation of the quark condensates in our next paper. 
 
Now let us proceed to the determination of  the heavy baryon masses. 
The  mass of  the heavy baryon can be obtained by (\ref{mass}).
In order to distinguish the mass split for the doublet(3/2,5/2), 
we expand the masses as 
\begin{equation}
\Lambda_\pm =\Lambda_0+\frac{\Lambda_K}{2M_h}+a_\pm\frac{\Lambda_M}{2M_h}
\end{equation}

The sum rules for $\Lambda_0$ $\Lambda_K$ and $\Lambda_M$ are shown 
in Fig.3 for the case  (\ref{I}),  where we use the 
threshold $s=2GeV$  
 and the gluonic condensate $\langle\alpha_sG^2\rangle
= 0.07GeV^4$\cite{reind}\cite{nar1}.  Fig.4 shows the
dependence of the $\Lambda_0$ on $s$, 
where it is shown that $s<\Lambda_0$ when 
$s=1.7GeV$. The $\Lambda_0$ is not very sensitive to $s$ around 
$s=1.8GeV$. It goes down  when  $\langle\alpha_sG^2\rangle$ 
becomes smaller.  
The sum rule for (\ref{I1}) is very similar to the case  (\ref{I}). 
It gives the same mass prediction  
within the uncertainty of these two  sum rules. The sum rule 
of (\ref{I3}) is much different from that of (\ref{I}) and (\ref{I1}). 
Fig.5 shows that the kinetic term $\Lambda_K$ is negative 
when $s\geq 1.6GeV$. We do not find find a stable region for the 
threshold $s$. Fig.6 shows the sum rule of (\ref{I3}) at 
$s=1.4GeV$. The situation of  (\ref{I4}) is similar to (\ref{I3}). 
Thus we obtain 
\begin{equation}
\Lambda_0 =1.8\pm 0.2GeV ,~~ \Lambda_K=0.5\pm0.5GeV^2  ,~~ 
\Lambda_M=0.05\pm0.05GeV^2.  
\end{equation}
for the sum rules (\ref{I}) and (\ref{I1}) and 
\begin{equation}  
\Lambda_0=1.2\pm0.2 GeV, \Lambda_K=0.06\pm0.05GeV^2, 
\Lambda_M=0.03\pm0.005GeV^2 
\end{equation}
for the sum rules (\ref{I3}) and (\ref{I4}). The error bars are  
from varying the parameters in a proper region.  
$\Lambda_M$ is very sensitive to $\langle\alpha_sG^2\rangle$. 
When $\langle\alpha_sG^2\rangle=0.04GeV$, it becomes  
unbelievably small. This is similar to the case in \cite{dai}. 
The reason may be that 
the mass split mainly is from the internal gluon exchange which we
 did not consider here. 
 $\Lambda_K/1GeV$ is much smaller that the masses 
of $b$ and $c$ quarks. Therefore the expansion of $1/M_h$  
stands well.   
Using the  quark masses \cite{grozin}\cite{dai}
\begin{equation}
m_b=4.8GeV,~~~~~~~~~~~m_c=1.4GeV,
\end{equation}
we predict the masses of the baryon doublet
(3/2,5/2) are
\begin{equation}
M_b=6.6\pm0.2GeV,~~~~~~~~~~~M_c=3.4\pm0.4GeV.
\end{equation}
for  (\ref{I}) and (\ref{I1}) and 
\begin{equation}
M_b=6.0\pm0.2GeV,~~~~~~~~~~~M_c=2.8\pm0.3GeV.
\end{equation}
for  (\ref{I3}) and (\ref{I4}).

Finally, let us give a brief summary. 
We have discussed the baryonic  currents 
 with  arbitrary quantum numbers  as well as 
their one-loop renormalization. 
Using the obtained currents, we have analyzed 
their two-point correlators. 
For a special case, we did the QCD sum rule for the masses of 
the heavy baryon doublet (3/2,5/2) up to  the order $1/M_h$. 
 Because we did not take account of the light quark mass and the 
quark condensates, we could not distinguish the two-point 
correlators  of the currents with different parity.
 Also because we did not take account of the internal gluon 
exchange, the mass split of the heavy baryon doublet is too small. 
We will include these discussions in our next paper.
   
\vspace{1.0cm}
{\bf Acknowledgment}
We would like to thank  A. A. Pivovarov and  S. Groote for
very useful discussions. The work of H.Y. J. is supported  
by the  Alexander von Humboldt.

\newpage
\par
{\huge\bf Figure captions}\\
\par
Fig.1 Feynman diagrams for the renormalization of the baryonic 
currents.\\

Fig.2  Feynman diagrams for the two-point correlator. The smaller 
dots stand for the current vertices, the bigger dots stand for 
the $1/M_h$-order interaction vertices.\\ 

Fig.3 The Sum rule for (\ref{I}).
Dashed line gives $\Lambda_0$ versus Borel variable $M$, 
solid line is $\Lambda_K/1GeV$ versus Borel variable $M$
and  dotted line is $\Lambda_M/1GeV$ versus Borel variable $M$.\\

Fig.4 The dependence of $\Lambda_0$ on $s$. 
Dashed line gives the result for $s=1.8GeV$,
solid line is for $s=1.7 GeV$ and  dotted line is for $s=2GeV$.

Fig.5 The dependence of $\Lambda_K/1GeV$ on $s$ for (\ref{I3}). 
Dashed line gives the result for $s=1.4GeV$,
solid line is for $s=1.2 GeV$ and  dotted line is for $s=1.6GeV$. 

Fig.6 The Sum rule for (\ref{I3}).
Dashed line gives $\Lambda_0$ versus Borel variable $M$,
solid line is $\Lambda_K/1GeV$ versus Borel variable $M$
and  dotted line $\Lambda_M/1GeV$ versus Borel variable $M$.\\ 

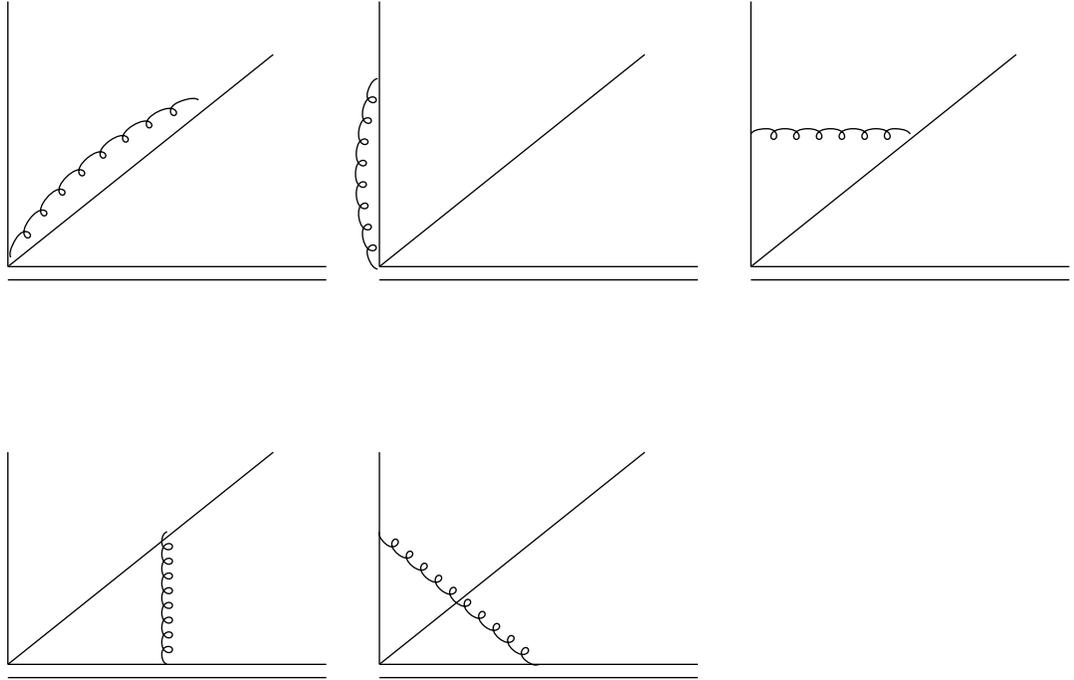
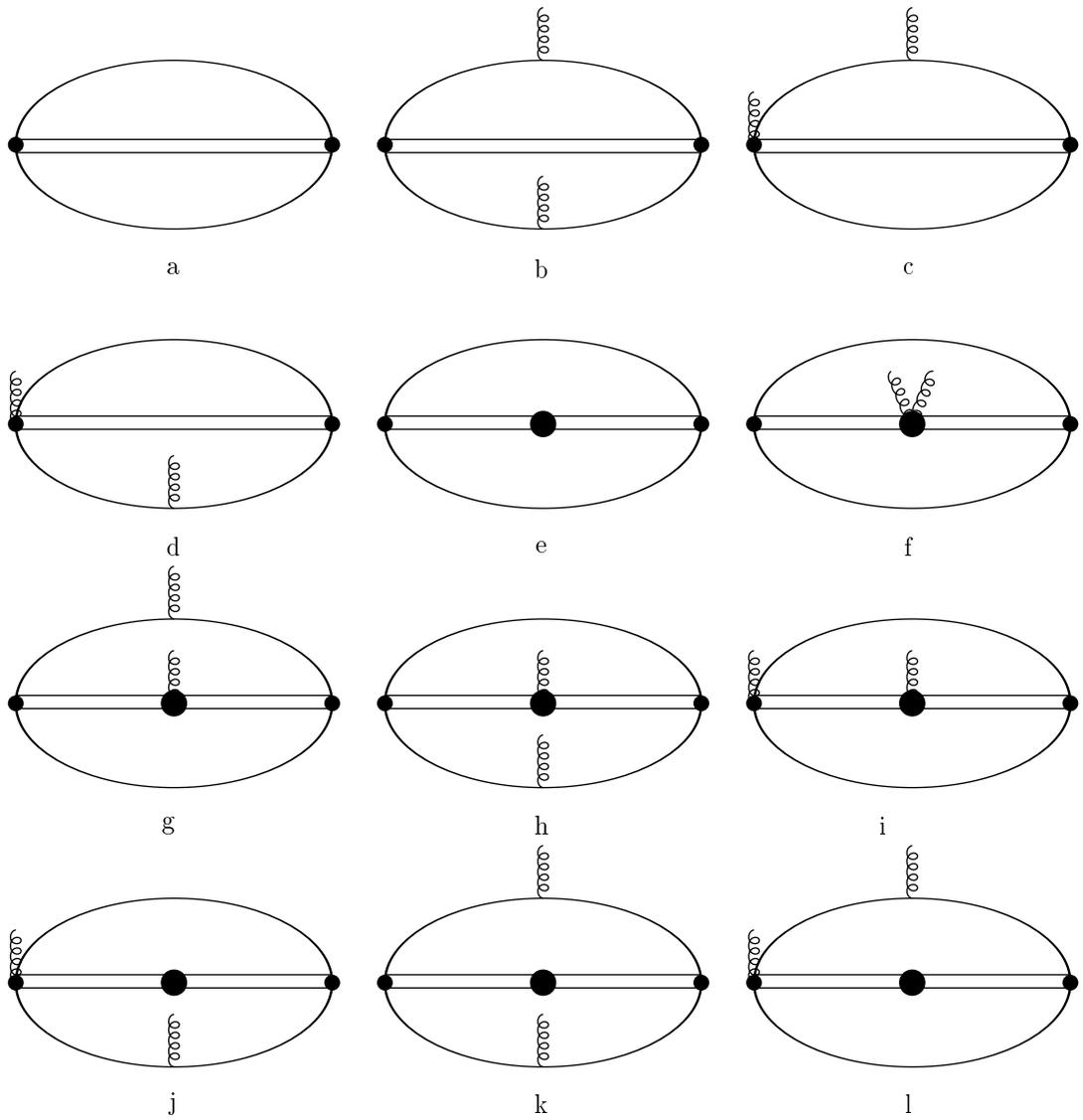
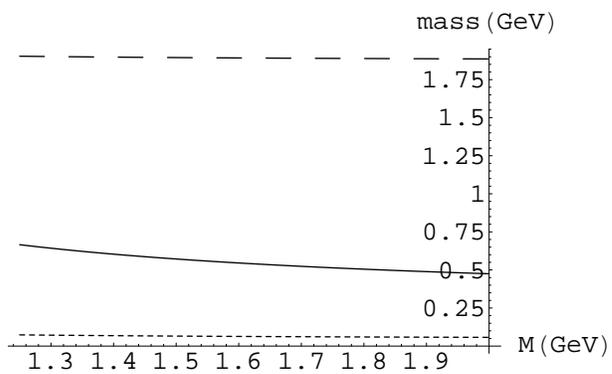
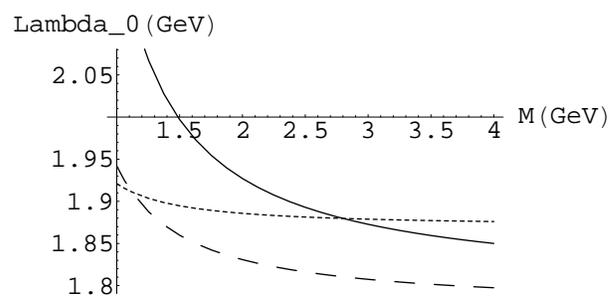
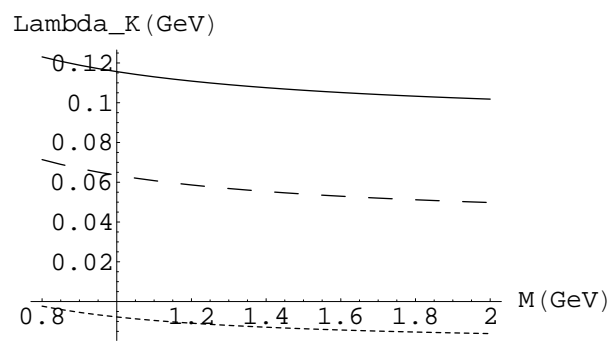
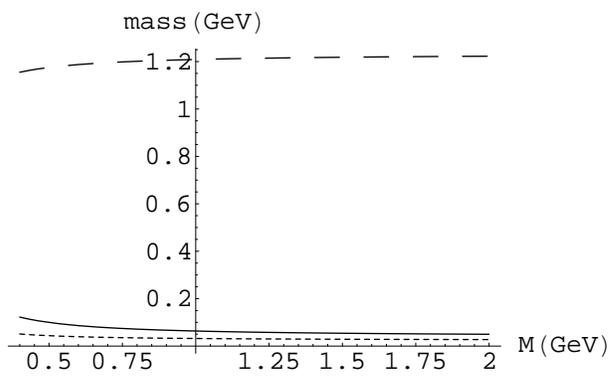
\end{document}